\documentclass[twocolumn,groupaddress]{revtex4}
\usepackage{color}
\usepackage{ulem}
\newcommand{\rev}[1]{#1}
\renewcommand{\em}{\it}
\renewcommand{\sout}[1]{}
\newlength{\swidth}
\newlength{\dwidth}
\usepackage{graphicx}
\usepackage{epsfig}
\begin{document}

\title{ A New Mechanism of Model Membrane Fusion Determined from
 Monte Carlo Simulation}

\author{M.\ M\"{u}ller}
 \affiliation{Institut f{\"u}r Physik, WA 331, Johannes Gutenberg
 Universit{\"a}t, D-55099 Mainz, Germany}

\author{K.\ Katsov}
\author{M.\ Schick}
\affiliation{Department of Physics, University of Washington,  Box
  351560, Seattle, WA 98195-1560} 

\date{\today}

\begin{abstract} 
We have carried out extensive Monte Carlo simulations of
the fusion of tense apposed bilayers formed by amphiphilic molecules
within the framework of a coarse grained lattice model. The fusion pathway
differs from the usual stalk mechanism. Stalks do form between the apposed
bilayers, but rather than expand radially to form an axial-symmetric
hemifusion diaphragm of the trans leaves of both bilayers, they promote in
their vicinity the nucleation of small holes in the bilayers.  Two
subsequent paths are observed:  (i) The stalk encircles a hole in one
bilayer creating a diaphragm comprised of both leaves of the other intact
bilayer, and which ruptures to complete the fusion pore. (ii) Before the
stalk can encircle a hole in one bilayer, a second hole forms in the
other bilayer, and the stalk aligns and encircles them both to complete
the fusion pore. Both pathways give rise to mixing between
the cis and trans leaves of the bilayer and allow for \rev{transient} leakage.
\end{abstract} \maketitle

\section{ Introduction }
Although membrane fusion is a fundamental biological process of
importance in fertilization, synaptic release, intracellular traffic, 
and viral infection, its
basic mechanism is not well understood. Much of the literature has focused
on fusion proteins whose function is, {\em inter alia}, to overcome the
energetic cost of bringing the bilayers to be fused to within a small
distance of one another, a step which places the membranes under tension
\cite{chen01}. There is accumulating evidence, however, that the subsequent
stages in the fusion pathway, the interruption of the integrity of the bilayers
and the molecular rearrangements that lead to the formation of the fusion pore
itself, are essentially lipidic in nature \citep{Zimmerberg99,Lentz00}.
A consequence of this view is that the fusion process can be studied,
both experimentally and theoretically, utilizing simple model membrane systems. 
Knowledge of the fusion mechanism in these simpler systems would illuminate 
additional roles that the proteins need to play in biological fusion.

\rev{The theoretical treatment of membrane fusion has, almost without
exception, been restricted to phenomenological models which decribe the
bilayer, not in terms of the  microscopic architecture of its components, 
but rather in terms
of the macroscopic  elastic properties of its monolayers.}
The common assumption is that these elastic moduli are uniform and
independent of membrane deformations \cite{Safran94}.  Although attractive
mathematically, this approach has its limitations.  For instance, it is
not clear whether the expansion of the membrane free energy to second
order in deformations is sufficient to describe the highly curved
intermediate structures, which may be involved in fusion.  Additional
approximations must be introduced to calculate the properties of junctions
of bilayers, which are not well described by simple bending deformations.
The energy of these structures has proven to be particularly sensitive to the
approximation used in their description
\cite{Siegel93,Kuzmin01,Kozlovsky02, Markin02}.  Importantly, application
of these approaches requires one to {\em assume} a particular fusion
pathway. The only pathways considered to date has been limited to variants
of one hypothesis \cite{Markin83,Kozlov83,Chernomordik85}. One starts with
two bilayers in close apposition. Lipids in the facing, or {\em cis},
layers rearrange locally and bridge the aqueous gap between the bilayers.
This results in the formation of an axially symmetric stalk. In most
versions, the stalk then expands radially and the {\em cis} layers recede.  
The {\em trans} layers make contact and produce an axially symmetric
hemifusion diaphragm. Nucleation of a hole in this diaphragm completes the
formation of an axially symmetric fusion pore. Because of the evolution of
the stalk into a hemifusion diaphragm in this model, we shall refer to it
as the ``hemifusion mechanism." \rev{Because only variants of the hemifusion
mechanism have been examined, and because the theory is
phenomenological, one does not know {\em a priori} in what systems this
pathway may be the most favored, or under what conditions. 
Some insight is gained by comparison with experiment which 
shows this hypothesized mechanism to be
consistent with a wide range of experimental observations of biological lipids
\cite{Monck96,Zimmerberg99,Jahn02}. However 
there is no direct evidence to confirm that this particular pathway is
that taken either by biological or laboratory-prepared model membranes}.

\rev{In light of the above, it would certainly be desirable to examine the
fusion pathway in a system whose components are described by a
microscopic model. Such examination has begun recently.}
A minimal model, consisting of rigid amphiphiles
\rev{of one hydrophilic and two hydrophobic segments}, and with no explicit
solvent, was studied with Brownian
dynamics simulations \cite{Noguchi01}.
At the same time, a model of more
complex, flexible, chain molecules, widely employed in the polymer
community, was used by us to study bilayers composed of amphiphilic,
diblock copolymers in a hydrophilic solvent \cite{Mueller02}. Such
copolymers are known, in fact, to form bilayer vesicles
\rev{which can undergo fusion} \cite{Discher99}.
This system was studied by means of Monte Carlo simulations.  Both
theoretical studies observed the formation of the initial stalk, but found
that the subsequent fusion pathway was {\em not} the usual hemifusion
mechanism, but involved intermediates that {\em broke the axial symmetry}.  
In particular, off to the side of the initial stalk, the formation of
small pores in each of the fusing bilayers was clearly seen. (We shall
refer to these small pores which span one bilayer only as ``holes" to
avoid confusion with fusion pores which span both bilayers.) \rev{ It is
intriguing that the two studies observed the same fusion pathway
even though the architecture of the constituents of the two systems
differed considerably, sharing little other than the generic property of
being amphiphilic and capable of bilayer self assembly.} \sout{As the
simulated microscopic models are very different, sharing only the
amphiphilic property of the constituents, the observed mechanism is a
candidate for explaining simple, generic fusion.}

The two investigations gave a first glimpse of a fusion pathway which
differs from the
hemifusion mechanism, but did not provide a great deal of quantitative
detail.  In this paper we present an extensive study of the same
microscopic model we employed previously, and offer sufficient
quantitative evidence to substantiate our earlier observations.

\sout{We conclude by observing that the mechanism we propose accounts for much
experimental evidence, in particular that the fusion process is leaky, the
fusion rate depends on lipid architecture and membrane tension, there is
mixing of lipids in the cis leaves before mixing of contents, and there is
also mixing of lipids between cis and trans layers.  In contrast to our
mechanism, the standard hemifusion mechanism is incompatible with the
first and last of these observations.}

\rev{Naturally we are concerned with  the question of whether the fusion
pathway we observe in our model system is relevant to membrane fusion in
biological systems. The architecture of the components in our system
obviously differs greatly from those of biological lipids, and it is not
clear how one should compare the systems. We make such an attempt
by calculating several
dimensionless ratios which can be formed from membrane parameters and
comparing those in our system with ratios characteristic of vesicles
formed of block copolymers, and of liposomes. (See Table I below.) 
Ultimately we cannot be sure of the systems to which our results
apply and under what conditions, save the very particular ones
that we have simulated for the particular case of block copolymers. In
this sense, our results must be evaluated in the same way as those from
the phenomenological theories; they must be compared to experiment. 
We do so in the Discussion. In particular we note that our mechanism
predicts that 
 the
fusion rate depends on lipid architecture and membrane tension, that there is
mixing of lipids in the cis leaves before mixing of contents, and that there is
{\em also} mixing of lipids between cis and trans layers. Of most
interest, our mechanism predicts that 
transient leakage is causally linked to the process of membrane fusion.}

\section{ Simulation details } 

Simulation of membrane fusion in a fully chemically realistic model would
be most valuable, because it could provide information about specific
structural changes on the atomic level. This would be particularly
important if changes in molecular conformations entailed a qualitative
spatial redistribution of hydrophilic and hydrophobic segments.
Unfortunately, the simulation of atomistically faithful models can only
follow the time evolution of a few hundreds of lipid molecules over a few
nanoseconds even on state-of-the-art supercomputers. Given that the time
scale of membrane fusion is on the order of milliseconds and involves
lengths on the order of a few tens of nanometers, an atomistic simulation
of the fusion process is not yet feasible and one has to resort to
coarse-grained models.

Coarse-grained models of amphiphilic chain molecules have been used with
great success to investigate \rev{common} \sout{the universal} 
features of self-assembly. Such
models retain only those molecular properties that are necessary for
self-assembly, such as the connectivity of hydrophilic and hydrophobic
portions along the amphiphilic molecule, and the mutual repulsion between
these different kinds of segments, and ignore specific chemical or
electrostatic interactions. The usefulness of this approach rests on the
observation that chemically very different systems, such as biological
lipids in aqueous environment and block copolymers in a homopolymer
environment, exhibit a common phase behavior and similar structural
patterns on length scales comparable or larger, than the molecular size.
The self-assembly of amphiphiles into bilayer membranes itself is an
example of a universal behavior, \rev{ {\em i.e.} one which does not 
depend on fine details of the underlying architecture.} It
has been successfully studied by
coarse-grained models \cite{Shillcock02}. \rev{We expect that all membranes
can be caused to fuse, however there may be several different pathways
which are taken by different systems under different conditions. Our
purpose here is to demonstrate one path which is taken in a system modeled
microscopically.}\sout{We expect the fusion of
membranes to exhibit a similar degree of universality, an expectation,
which underlies every theoretical study of membrane fusion to date. In
this context, coarse-grained models can provide valuable information about
the salient features of fusion in model membranes, as will be demonstrated
below.}

We employ the bond fluctuation model \cite{Carmesin88} of a polymer chain,
which has been used previously to study pore nucleation in a symmetric
bilayer membrane under tension \cite{Mueller96}. Much is known about the
structure and thermodynamics of this model, and the parameters can be
mapped onto the standard Gaussian chain model of a dense mixture of
extended molecules. In this three-dimensional lattice model, each segment
occupies a lattice cube. No two occupied cubes can share any corner, a
rule that mimics hard-core repulsion interaction. \rev{Furthermore this
ensures that the lattice spacing is sufficiently smaller than the width
of interfaces so that the effect of the discretization of space is 
minimal.} To ensure that the chain
of segments can not intersect itself, the segments are connected by bond
lengths that cannot be too large. In particular, neighboring segments
along the chain can be connected by one of 108 bond vectors of lengths
$2,\sqrt{5},\sqrt{6},3$ or $\sqrt{10}$ measured in units of the lattice
spacing $u$. \rev{The angles between adjacent monomers can take on any of 
87 values.} The large number of bond vectors and the extended segment
shape allow a rather faithful approximation of continuous space, while
retaining the computational advantages of lattice models. The amphiphilic
molecules consist of $11$ hydrophilic segments and $21$ hydrophobic
segments. This asymmetry mimics the ratio of head and tail size in
biologically relevant lipid molecules, and is slightly smaller than
employed by us previously \cite{Mueller02}. We reduced, in this study, the
asymmetry of the molecules so that a solvent-free system not only would be
in a lamellar phase ($L_\alpha$), but would also be further than in our
previous study from the boundary separating the lamellar and
inverted-hexagonal ($H_{II}$) phases.  
The solvent in our system is
represented by chains of $32$ hydrophilic segments, {\em i.e.} we conceive
a hydrophilic chain as a small cluster of solvent molecules, \rev{just as in
other coarse-grained modeling \cite{shelley}} The mean
head-to-tail distance of the amphiphiles and solvent molecules is $17u$.
Like segments attract each other and unlike segments repel each other via
a square well potential which comprises the nearest $54$ lattice sites.
Each contact changes the energy by an amount $\epsilon=0.17689k_BT$. This
corresponds to an intermediate segregation $\chi N \approx 30$ in terms of
the Flory--Huggins parameter $\chi$.  If we increased the incompatibility
much more, we would reduce the interfacial width between hydrophilic and
hydrophobic segments to the order of the lattice spacing and the local
structure of the lattice model would become important. If we decreased the
incompatibility, we would reduce the clear segregation between hydrophilic
and hydrophobic regions. Similarly if we replaced the solvent homopolymers
by monomers, we would effectively reduce the incompatibility
\cite{Matsen95}, \rev {and again
reduce the segregation between the diblock and solvent hydrophilic
segments and the diblock hydrophobic segments. Were we to increase the
incompatibility to restore the desired degree of segregation, we would
again reduce the interfacial width of the membrane to an extent that
lattice effects would become important.}

Monte Carlo simulations are performed in the canonical ensemble, except
for some runs described in section III. The segment number density, {\em
i.e.} the fraction of lattice cubes occupied by segments, is fixed at
$\rho=1/16$. The conformations are updated by local segments displacements
and slithering-snake-like movements. The different moves are applied with
a ratio $1:3$. We count one attempted local displacement per segment and
three slithering-snake-like attempts per molecule as four Monte Carlo
steps (MCS).  This scheme relaxes the molecular conformation rather
efficiently. The latter moves do not mimic a realistic dynamics of lipid
molecules and we cannot identify straightforwardly the number of Monte
Carlo steps with time. The density of hydrophilic and hydrophobic
segments, however, {\em is conserved} so that the molecules move
diffusively.  Moreover, the molecules cannot cross each other during their
diffusive motion. In this sense we have a slightly more realistic time
evolution on local length scales than in dissipative particle dynamics
simulations \cite{Shillcock02}, but Monte Carlo simulations cannot include
hydrodynamic flows, which might become important on large length scales.  
At any rate, we do not expect the time sequence to differ qualitatively
from that of a simulation with a more realistic dynamics on time scales
much larger than a single Monte Carlo step. Most importantly, fusion is
thought to be an activated process, therefore the details of the dynamics
only set the absolute time scale, but the rate of fusion is dominated by
free energy barriers encountered along the fusion pathway, which are
independent of the actual dynamics used.

\section{ Preparation and properties of a single bilayer }

\begin{figure*}[ht]
\includegraphics[width=\dwidth]{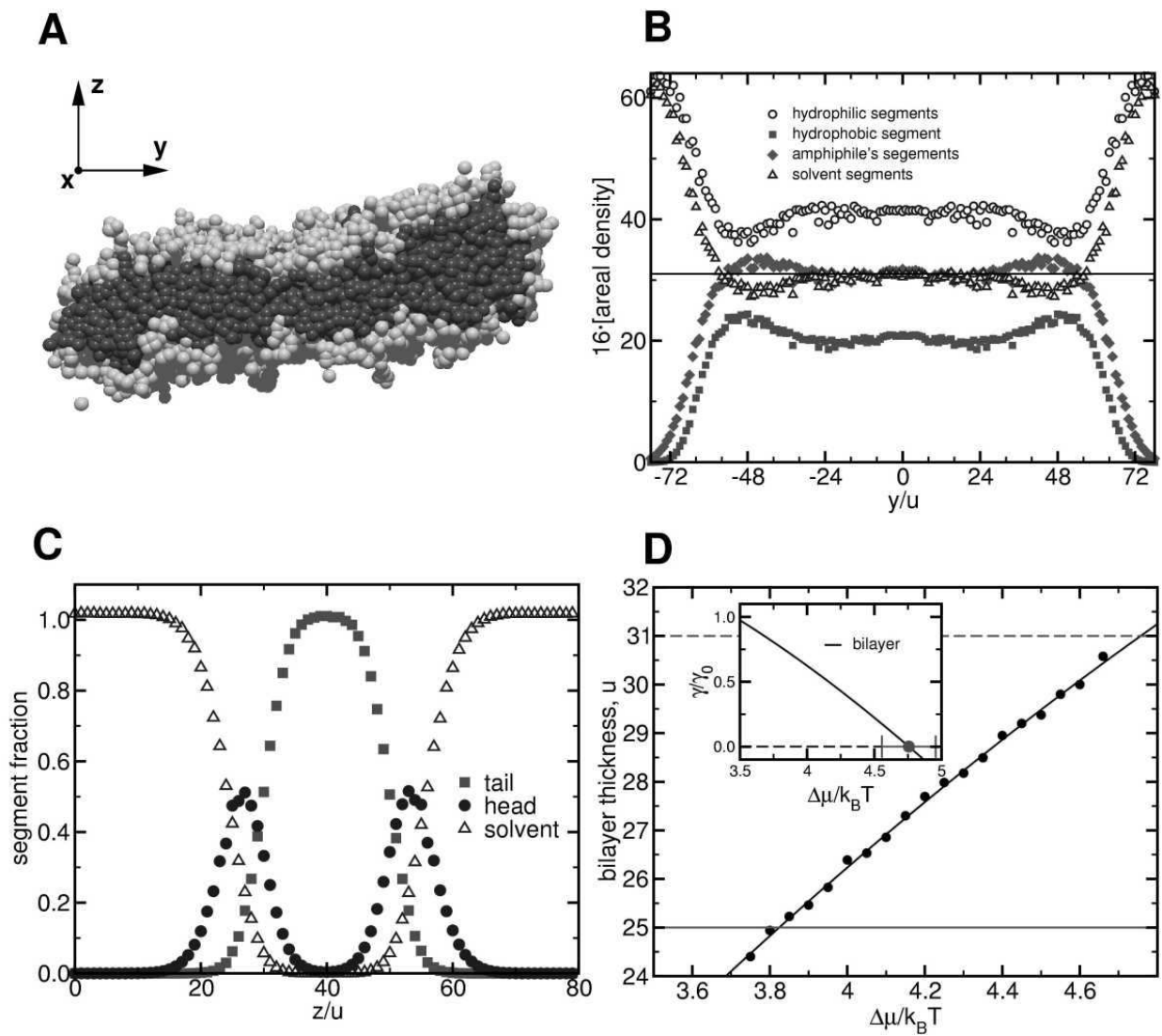}
\caption{{\bf (a)} Snapshot of an isolated bilayer in the
tensionless state. Hydrophobic and hydrophilic segments of amphiphiles 
are shown as dark and light gray spheres. Solvent segments are not shown
for clarity.
{\bf (b)} Density profiles along the $y$ axis. The edge of
the bilayer is thicker than its middle. {\bf (c)} Profiles across the
bilayer for a lateral patch of size $40u \times 40u$. {\bf (d)} Dependence
of the bilayer thickness on the exchange chemical potential $\Delta \mu$
between amphiphiles and solvent. The inset displays the tension $\gamma$
of the bilayer as a function of exchange potential.}
\label{fig:1-bilayer}
\end{figure*}

It seems clear that bilayers that are under no stress will not undergo
fusion, as there is no free energy to be gained by doing so. So to promote
fusion, we have subjected the studied bilayers to lateral tension. This
has been done by providing the system with fewer molecules than are needed
to span the given area of our sample cell with bilayers that are
tensionless. Of course we need to know just how many molecules are needed
to make a tensionless bilayer that spans the cell. To determine properties
of the tensionless bilayer,  \rev{we made use of the definition of 
the tension in
this liquid-like bilayer as the
derivative of the free energy with respect to the bilayer area 
at constant temperature and particle number.} 
We therefore investigated an isolated bilayer with a straight,
free edge. A simulation cell of size $64u\times 200u\times 64u$ with
periodic boundary conditions in all dimensions was used. The bilayer,
oriented in the $x-y$ plane, spanned the system in the short, $x$,
direction, but did not span the system in the long, $y$, direction. Its
extension in this direction adjusted itself until it neither grew nor 
shrank. Thus the surface tension,
$\gamma$, of the bilayer was zero. \rev{This vanishing value includes, of
course, the contributions from the fluctuations of the bilayer. 
Even though the tension vanished,}
these fluctuations of the mid-plane were not very large due to the stiffness of
the rather small patch of membrane considered.  A typical snapshot of the
bilayer configuration is shown in Fig.~\ref{fig:1-bilayer}{\bf (a)}.  
Rearrangement of amphiphiles at the bilayer free edges is clear. 
The average profile along the
$y$ axis, the long axis of the bilayer, is presented in panel {\bf (b)}.
To obtain it, we have averaged the profiles along the $x$ and $z$
direction and estimated the instantaneous angle the bilayer makes with the
$z$ direction (to correct for the difference between projected and true
area). We observe for these
laterally averaged profiles that the edge of the bilayer is slightly
thicker than the middle, the increase is about $7\%$ for the amphiphilic
segment density and about $16\%$ for the density of hydrophobic segments.
\rev{Away from the edge, the densities decay exponentially to those  
of the uniform bilayer
({\em i.e} without an edge)}, and we  estimate the thickness of the
tensionless bilayer from that in the middle, finding it to be $d_0=31u$. 

The profiles across a single bilayer of thickness $d_0=31$ are shown in
Fig. \ref{fig:1-bilayer}{\bf (c)}. They were obtained by simulation in a
cell $40u\times 40u\times 80u$ in which the bilayer spanned both short
directions. One sees that hydrophobic and hydrophilic regions are clearly
separated, but there is some interdigitation of the hydrophobic tails
emerging from the opposing monolayers.

\rev{Knowing the thickness of the tensionless bilayer, we know the number of
molecules  needed to span the simulation cell with such a bilayer, and
can control tension by varying the number of molecules introduced into 
the cell. 
 We cannot determine this tension, as one might in a Molecular
Dynamics simulation, from the excess tangential stress in the
interfacial zone  because we employ a lattice model. Nevertheless, we can
determine the tension purely from thermodynamic relations. To
do so, we} 
assembled a single bilayer in a system of size $156u \times 156u \times
64u$, where the bilayer spanned the system in the $x-y$ plane.  
Using semi-grand
canonical identity switches between amphiphiles and solvent, we controlled
the exchange potential $\Delta \mu$ between the species and monitored the
thickness of the bilayer (measured by the areal density of amphiphiles).
The dependence of the bilayer tension on the chemical potentials of the
amphiphile, $\mu_C$, and solvent, $\mu_A$, is given by the Gibbs
absorption isotherm \rev{\cite{davis}:} 
\begin{equation} L^2 {\rm d}\gamma = - \delta n_C
{\rm d}\mu_C - \delta n_A {\rm d}\mu_A \approx - \delta n_C {\rm d}\Delta \mu
\label{eqn:gibbs} \end{equation} 
where $\delta n_C$ and $\delta n_A$ are
the excess number of molecules in the bilayer. In the last approximation
we have assumed that the liquid is incompressible {\em i.e.} $\delta n_A
\approx - \delta n_C$, and the solubility of the amphiphile in the solvent
is vanishingly small. Results of the simulation for the number of
amphiphiles $\delta n_C$ as a function of the exchange potential $\Delta
\mu=\mu_C-\mu_A$ are shown in Fig.~\ref{fig:1-bilayer}{\bf (d)}. Using the
thickness of the tensionless bilayer, we can estimate the tension of an
arbitrary bilayer as a function of exchange potential or of thickness by
integrating Eq. (\ref{eqn:gibbs}).  The results, in reduced units of bare
A-B homopolymer interfacial tension $\gamma_0=0.068k_BT/u^2$,
are shown in the inset of
Fig.~\ref{fig:1-bilayer}{\bf (d)}.  Dashed lines in
Fig.~\ref{fig:1-bilayer}{\bf (d)} and the inset correspond to the
tensionless bilayer. Comparison of the relevant structural and elastic 
properties of the polymersomes, liposomes and simulated membranes 
is provided in Table~\ref{table1}.

We are now in a position to simulate bilayers under a
given tension in the canonical ensemble.  Knowing the area of our
simulation cell, and the segment density, we add the number of amphiphiles
which will produce a bilayer of a given thickness. From
Fig.~\ref{fig:1-bilayer}{\bf (d)}, we know what tension is placed on this
bilayer. For our study of two bilayers under tension, we have chosen their
thickness to be $d=25u$, smaller than the thickness $d_0=31u$ of the
tensionless bilayer. This corresponds to a tension of the order of $\gamma
/\gamma_0 \approx 0.75$ and an area expansion, $\Delta A/A_0\approx 0.19.$
 We know from our
simulations that a single bilayer of the thickness chosen, $d=25u$, is
metastable on the time scale of fusion, {\em i.e.} the small
holes, which appear transiently, do not grow past their critical size on
the time scale of fusion in our simulations. 

\section{ Preparation of a System of Two Bilayers }

We begin with a system containing only amphiphiles. It is $156u\times
156u\times 25u$ with periodic boundary conditions in the long directions,
and hard, impenetrable, walls in the short direction.  They attract the
hydrophilic portion of the amphiphile and repel the hydrophobic portion.
These interactions extend over two layers nearest to the wall, and each
contact changes the energy by $0.6 k_BT$. The amphiphiles assemble into a
bilayer structure which is free of defects.

\begin{figure*}[ht]
\includegraphics[width=\swidth]{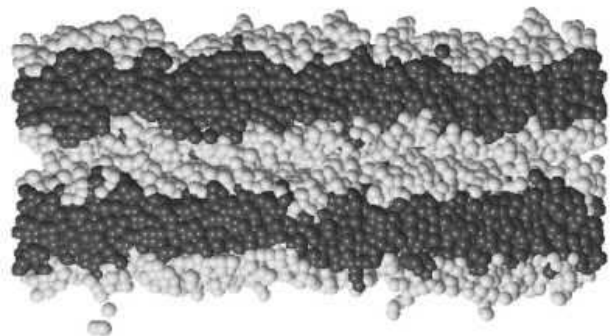}
\caption{Snapshot of the initial configuration in the two bilayer
  system. Hydrophobic and hydrophilic segments of amphiphiles are
  shown as dark and light gray spheres. Solvent segments are not shown for
  clarity.}
\label{fig:2-bilayer}
\end{figure*}

Two of these flat bilayers are then stacked on top of each other with a
distance of $\Delta$ between them, and are embedded into a simulation cell
with geometry $156u \times 156u \times 126u$.\rev{ There are no walls at this
point, and periodic boundary
conditions are utilized in all three dimensions. The conditions of flat
bilayers} mimic the
approach of two vesicles whose radii of curvature are much larger than the
patch of membrane needed for fusion. The solvent of homopolymers is then
inserted into the simulation cell via grand canonical, configurational
bias, Monte Carlo moves at infinitely large chemical potential of the
homopolymer until the segment number density of $\rho=1/16$ is reached.  
The initial distance $\Delta$ between the bilayers translates into the
thickness of the residual solvent layer between the two membranes. We have
carried out the most extensive series of runs with $\Delta=10u$ and unless
specified otherwise, all our results are for that separation. \rev{Because the 
solvent homopolymers are flexible coils, and each represents a cluster of 
solvent molecules, many layers of solvent segments are represented 
between the bilayers at this separation.} In our
previous simulations \cite{Mueller02}, we set $\Delta=0$ and observed
qualitatively similar behavior as we do with this larger separation.  We
increased the separation for this extensive study because, as expected,
the rate of fusion events decreased (see Sec.
V. below) and this allowed us to observe the sequence of structural
rearrangements more clearly than in our previous work. The separation
chosen, a bit less than half the thickness of one bilayer, is comparable
to the separation at which fusion occurs when mediated by hemagglutinin
\rev{\cite{Flint00}}.  A snapshot of the two
bilayers is shown in Fig.~\ref{fig:2-bilayer}.  Hydrophobic and
hydrophilic segments of amphiphiles are shown as dark and light gray
spheres. For clarity solvent segments, which are present in the
simulation, are not shown

\sout{Periodic boundary conditions are utilized in all three dimensions.}
Thirty-two independent starting configurations were prepared, each
containing $194,688$ segments corresponding to about $3,613$ amphiphiles
and $3,708$ solvent molecules. After every $25,000$ Monte Carlo steps, a
configuration was stored for further analysis. \rev{Ten thousand hours of
CPU time were utilized in the course of this investigation, with 
thirty two processors running for about two weeks.}

\section{ The Process of Fusion } 
\begin{figure*}[ht]
\includegraphics[width=\swidth]{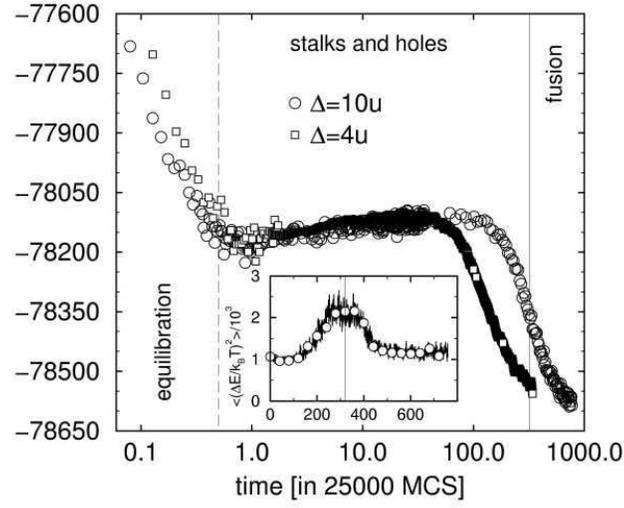}
\caption{
  Evolution of internal energy in fusion simulations. The two
  curves correspond to initial bilayer separations $\Delta=4u$
  (squares) and $\Delta=10u$ (circles). To reduce fluctuations,
  the data are averaged over all 32 configurations at equal time 
  and additionally over small time windows. The large negative value
  of the energy mirrors the attractive interactions in the solvent.
  The inset shows the sample-to-sample energy fluctuations as a function of time.
  Large fluctuations identify the onset of fusion.
}
\label{fig:energy}
\end{figure*}

It is straightforward to monitor the
internal energy of the system during the simulation because this energy
arises solely from contacts between segments, and the locations of all
segments are known. (In contrast, the {\em free} energy cannot be obtained
directly.) We show in Fig.~\ref{fig:energy} the behavior of the internal
energy of two systems, one separated by a distance $\Delta=4 u$ (squares),
and the other with $\Delta=10 u$ (circles). The energy is plotted in units
of $k_BT$, as a function of time, in units of $25,000$ Monte Carlo steps.  
The energy initially decays, which reflects the equilibration of the
system.  During this initial relaxation of the starting configuration the
interface between the bilayer and the solvent adjusts locally.  The time scale
of this initial relaxation (less than $25,000$MCS)  is independent of
the distance between the bilayers, and is about two orders of
magnitude smaller than the time scale on which the fusion pore forms. Due to
this separation of time scales between initial relaxation and fusion we do
not expect the preparation of the starting configuration to affect the
fusion process. \rev{Similarly we do not expect our results to
depend on our particular choice of relaxation moves, as other choices
would also lead to relaxation of the bilayers which 
takes place on a much shorter time scale than does fusion.}

After the initial relaxation, two subsequent time regimes can be
identified.  First the energy rises slowly. Two mechanisms contribute to
this increase of the energy. On the one hand, capillary waves of the
hydrophilic/hydrophobic interfaces become thermally excited. They increase
the effective interface area and thereby lead to a slow increase of the
energy. Additionally, undulations result in the formation of stalks and
holes. We shall discuss the details of this process in the next
subsection. Later, around $320 \times 25 000$ MCS the energy decreases
rapidly.  This final decrease of the energy results from the fusion of the
membranes which releases some of the tension stored in them. As noted
above, the fusion occurs more rapidly the closer the bilayers, as
expected. The increase in energy preceding fusion reflects the formation
of fusion intermediates, the focus of our study.

The inset of Fig.~\ref{fig:energy} shows the fluctuations in the energy,
{\em i.e.}, the fluctuations between the 32 different runs at equal time.
Strong fluctuations indicate energy differences between the independent
runs. The peak at around $t \approx 320$ indicates that some systems have
already formed a fusion pore (and therefore have a lower energy) while
others systems have only stalks and holes (and therefore have a higher
energy).  The vertical bar indicates the time we have chosen to indicate
on several figures the onset of fusion. The width of the peak provides an
estimate for the spread of the time at which a fusion pore appears.

\subsection{The stalk and associated hole formation}

\begin{figure*}[ht]
\includegraphics[width=\swidth]{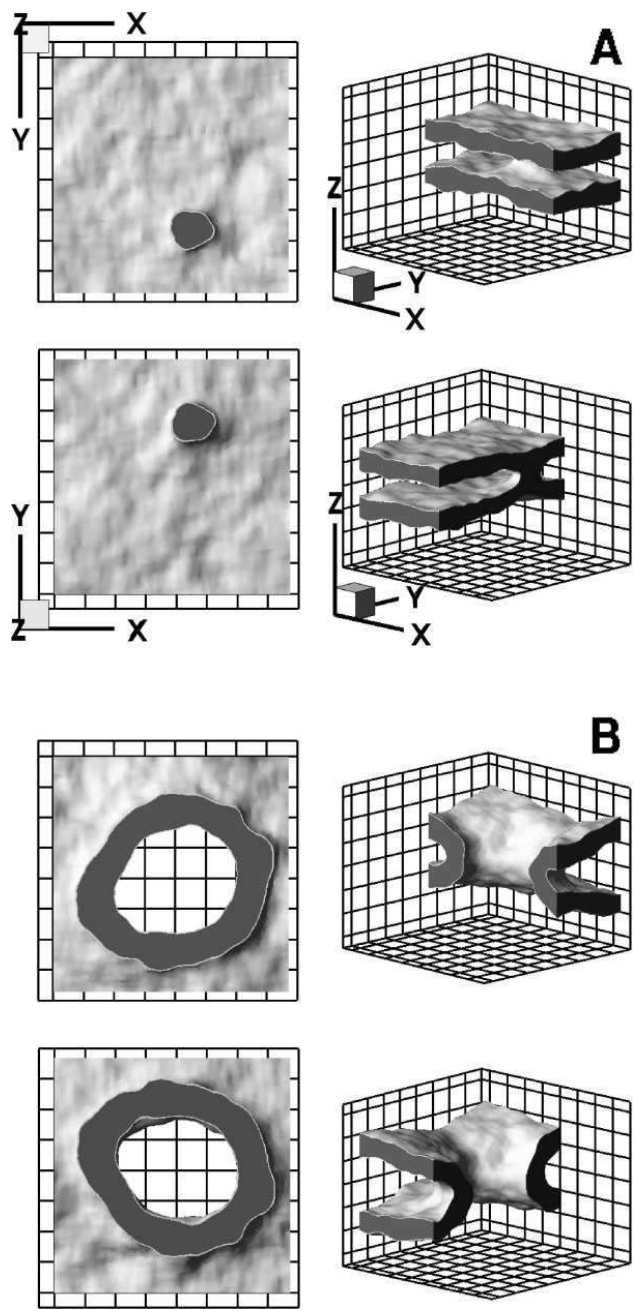}
\caption{Representative snapshots of {\bf (a)} the stalk intermediate and
{\bf (b)} the complete fusion pore from one of the simulation runs.
Each configuration is shown from four different viewpoints.
Hydrophobic core is shown as dark gray, the hydrophilic--hydrophobic
interface (defined as a surface on which densities of hydrophilic and
hydrophobic segments are equal)
is light gray. Hydrophilic segments are not shown for clarity.
Top- and bottom- left sub-panels have been generated by cutting 
the system along the middle $x-y$-plane, the top and bottom
halves are viewed in the positive (up) and negative (down) $z$-direction
correspondingly. Top- and bottom- right sub-panels are side views with
cuts made by $x-z$ and $y-z$ planes correspondingly. Grid spacing is
$20u$. 3D-orientation axis are the same for all snapshots and 
shown in panel {\bf (a)}.
}
\label{fig:mechanism1}
\end{figure*}

\begin{figure*}[ht]
\includegraphics[width=\swidth]{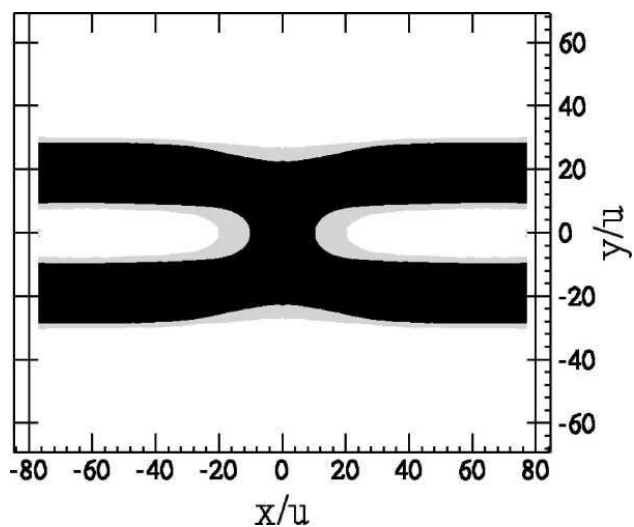}
\caption{Density distribution of segments in the
         stalk, averaged over all simulation runs. At each point only
         the majority component is shown: solvent as white, hydrophobic
         and hydrophilic segments of amphiphiles as black and gray
         respectively.}
\label{fig:stalk_profile}
\end{figure*}

During the initial stage of simulations the fluctuating bilayers collide
with one another frequently and sometimes form small local
interconnections. For the most part, these contacts are fleeting.
Occasionally we observe sufficient rearrangement of the amphiphiles in
each bilayer to form a configuration, the stalk, which connects the two
bilayers, (see Fig.~\ref{fig:mechanism1}{\bf (a)}) and which is not as
transient.  Such a stalk was hypothesized long ago to be involved in the
initial stages of fusion \cite{Markin83,Kozlov83}.  In contrast to {\em
stable arrays} of stalks which have been observed in block copolymer melts
\cite{Disko93} and in lipid systems \cite{Yang02}, those we see are
isolated, and increase the free energy of the system. We infer the latter
from two observations: that the appearance of stalks is correlated with
the increase in the {\em internal energy} of the system as a function of
time shown in Fig.~\ref{fig:energy}; that some stalks vanish without
proceeding further to a fusion pore. Thus it appears that the stalk
represents a local minimum along the fusion pathway. Density profiles of
the hydrophilic and hydrophobic parts of the amphiphiles in the presence
of the stalk, and obtained by averaging over configurations, are shown in
Fig.~\ref{fig:stalk_profile}. The dimples in the membranes at each end of
the stalk axis are notable.  What can barely be seen is a slight thinning
of each bilayer a short distance from the axis of the stalk.

\begin{figure*}[ht]
\includegraphics[width=\swidth]{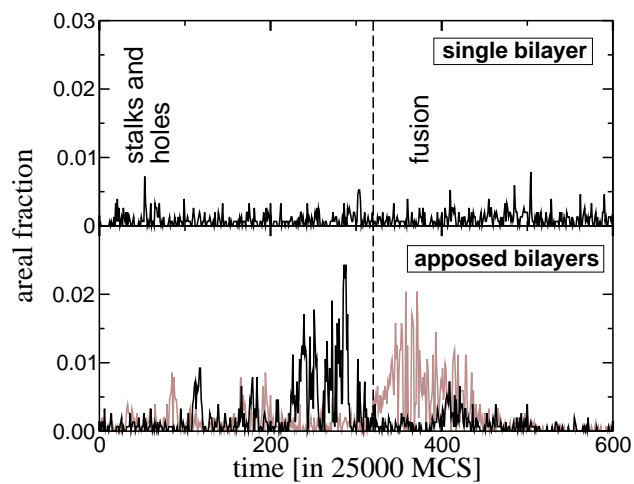}
\caption{Area of holes vs. time in the system of two apposed bilayers 
(gray for one bilayer and black for the other on the bottom panel) and in an isolated bilayer 
(top panel).}
\label{fig:holes_2vs1}
\end{figure*}

After stalks are formed, the rate of formation of holes in either of the
two bilayers goes up markedly. This can be seen in
Fig.~\ref{fig:holes_2vs1}. in which we plot the fractional area of holes
as a function of time for the system of two apposed bilayers, and compare
it to the rate of hole formation in an isolated bilayer. In
contrast to the large increase in the area of holes formed in the apposed
bilayers at ``time'' $t=200 \times 25 000$ MCS, the fractional area in
single bilayers fluctuates somewhat about an average value which is rather
constant over time at a value of approximately $0.004$. Comparison with
Fig.~\ref{fig:energy} shows that the increase in the rate of hole
production in the apposed bilayers in this system with bilayer spacing
$\Delta=10u$ is correlated in time with the decrease in the energy of the
system, and it is reasonable to infer that the decrease in energy is
caused by the production of holes and, later, the fusion pore. Similarly,
during the time before this increase in hole production, stalks are
forming, and it is also reasonable to infer that the increase in energy is
due to their formation.

The locations of stalks and holes are correlated; holes form close to the
stalks, and the stalk {\em elongates and moves} to surround the hole. A
snapshot of this is shown in Fig.~\ref{fig:mechanism2}{\bf (a)} and {\bf
(d)}.  In both snapshots an elongated stalk is seen and a small hole is
formed in the upper bilayer next to the stalk.  The extent to which holes
are, on average, found close to a stalk can be determined from the
hole-stalk correlation function
\begin{equation} 
g(r)\equiv \frac{\sum_{r_s,r_h}\delta(|{\bf r}_s-{\bf
r}_h|-r) P_{sh}({\bf r}_s,{\bf r}_h)}
                {\sum_{r_s,r_h}\delta(|{\bf r}_s-{\bf r}_h|-r)}
\end{equation}
where $P_{sh}({\bf r}_s,{\bf r}_h)$ is the joint probability that the
lateral position ${\bf r}_s$ is part of a stalk and ${\bf r}_h$ is part of
a hole, and $\delta(r)$ is the Dirac delta function. The value of $g(r)$
at large distances $r$ is proportional to the product of the areal
fraction of holes and stalks.  This correlation function is shown in
Fig.~\ref{fig:stalk_hole_corr}.  The scale of $g(r)$ increases with time
indicating the simultaneous formation of stalks and holes. The figure
shows that the correlation peaks at a distance of about $16u$, and falls
rapidly at larger distances. (Recall that each bilayer has an average
thickness of $25u$.)

It is not difficult to understand why the presence of a stalk promotes
hole formation. First, if the hole forms close to a stalk, then the line
tension, or energy per unit length $\lambda$, of that part of the hole
near the stalk is significantly reduced. This can be seen from the
schematic in Fig.~\ref{fig:stalk_hole_scheme}. 
In the upper part of the figure, we show a hole which
has formed far from a stalk, while in the lower, we show a hole which has
formed close to one. It seems clear that the line tension in the latter 
is reduced
simply due to the reduction of curvature of the hydrophobic-hydrophilic 
interface. 
The second
reason that the stalk formation encourages the appearance of holes is due to
the slight thinning of the membrane in the vicinity of the stalk to which
we alluded earlier. Further it has been suggested recently that the
local surface tension in the neighborhood of a defect, such as a stalk,
is increased significantly \cite{Kozlovsky002} 
making such a location the likely site of hole formation.

Now that one hole has formed next to the stalk, and the stalk has begun to
surround it, two other events occur to complete the formation of
the fusion pore. They are: i) a second hole forms in the other
bilayer, ii) the stalk surrounds the hole(s) to form the rim
of the fusion pore. We have observed these steps to occur in either order,
and will briefly discuss them separately.
   
\subsection{Pathway 1. Rim formation followed by appearance of a second
hole}
\begin{figure*}[ht]
\includegraphics[width=\dwidth]{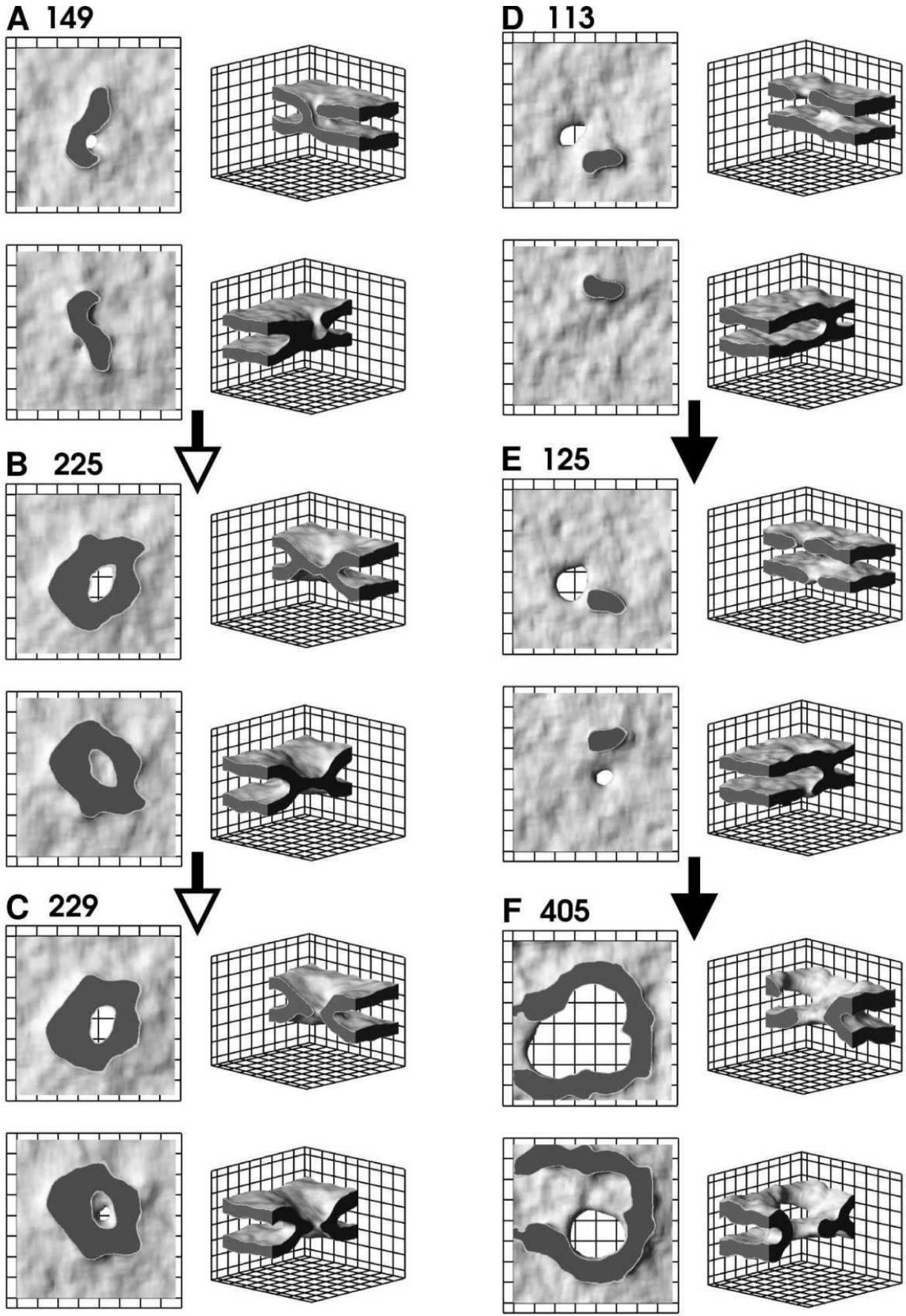}
\caption{Two observed pathways of fusion process.
The snapshots are taken from two representative simulation runs.
Each configuration is numbered by the time (in multiples of
$25,000MCS$) at which it was observed.
See Fig.~\ref{fig:mechanism1} for explanation of the graphics shown.
For discussion of the mechanism see text.}
\label{fig:mechanism2}
\end{figure*}

\begin{figure*}[ht]
\includegraphics[width=\swidth]{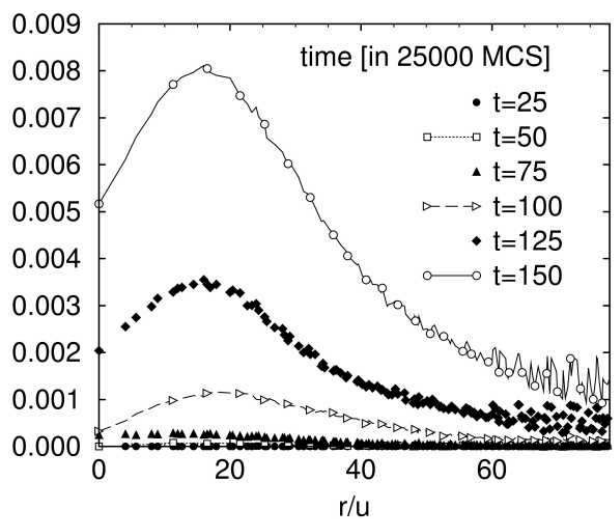}
\caption{The hole--stalk correlation function at early times.}
\label{fig:stalk_hole_corr}
\end{figure*}

\begin{figure*}[ht]
\includegraphics[width=\swidth]{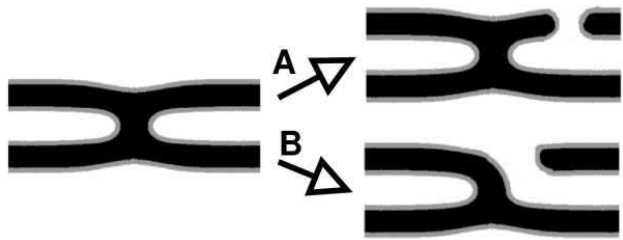}
\caption{Schematic explanation of the line tension reduction near the stalk.}
\label{fig:stalk_hole_scheme}
\end{figure*}

In this scenario,  a hole appears in one bilayer and the stalk
completely surrounds it rather rapidly. A snapshot of
the system  in this configuration is shown in
Fig.~\ref{fig:mechanism2}{\bf (b)}. This looks very
much like a hemifusion diaphragm which has been suggested by many authors
as an intermediate stage in fusion \cite{Markin83,Chernomordik85,Siegel93}. 
However, this diaphragm is quite different from the usual hemifusion
one that consists of two trans monolayers of the fusing membranes.
In contrast, the diaphragm we observe is made of one of the pre-existing
bilayers; that is, it is made  of {\em cis and trans} leaves.
The appearance of a
hole in this diaphragm, as shown in Fig.~\ref{fig:mechanism2}{\bf (c)}, 
and its expansion completes the formation of the fusion pore.

\subsection{Pathway 2. Appearance of second hole followed by rim formation}
In this scenario, a hole appears in one bilayer and, before the stalk
completely surrounds it, a second hole appears in the other bilayer. The
stalk tries to surround them both, and aligns them in doing so. 
In Fig.~\ref{fig:mechanism2}{\bf (e)} we show one stage in this process. 
One sees a large hole in the upper bilayer.
A small hole is formed in the lower bilayer
next to the stalk. Eventually, the stalk aligns and completely encircles the
holes (see Fig.~\ref{fig:mechanism2}{\bf (f)}) to form the 
final fusion pore shown in
Fig.~\ref{fig:mechanism1}{\bf (b)}. \rev{Again, the driving force for the stalk to
surround the two holes is the reduction in their (bare) line tension. 
Because the stalk aligns and
surrounds two holes,  we observe this pathway to be
somewhat slower than that of pathway 1 in which the stalk need only
surround one hole.}

Once the fusion pore has formed, by either of the above mechanisms, it 
expands, driven by the reduction in surface tension. The growth of 
the fusion pore eventually slows and ends as the pore reaches its
optimum size determined by the finite size of our cell.

\section{Discussion} We have carried out extensive Monte Carlo simulations
on the fusion of two bilayer membranes comprised of amphiphilic molecules
immersed in solvent. The amphiphiles and solvent are modeled by copolymers
and homopolymers respectively. The membranes are under tension. The
mechanism of fusion that we see begins with a stalk, as posited years ago,
and incorporated in almost all fusion scenarios.  However what follows
after stalk formation is different from  all other mechanisms
which have been proposed save that presented independently by Noguchi and 
Takasu \cite{Noguchi01}. In particular, the fusion intermediates we see
break the axial symmetry which has been assumed in almost all previous
calculations. \rev{We observed that the stalk destabilizes the
bilayers by catalyzing  the creation of small holes in them. We argued
that the mechanism behind this is quite simple: the energy per unit
length of the edge of a hole is reduced when the edge is adjacent to a
stalk. For the same reason, the stalk will try to surround the hole
formed in one bilayer once
the hole has appeared. Two slightly different pathways to the final fusion pore
were observed differing only on whether the hole in the second bilayer, 
which is necessary for complete fusion, appears before or after the
stalk completely surrounds the first hole.}   

\rev{The question now arises as to whether the pathway we see in the model
system is that which occurs in biological fusion. There are many  
differences between the model studied and a biological system. Perhaps 
the most obvious is that we have modeled flexible, single chain
 block copolymers, not lipids with two semiflexible tails and a rigid 
 head. How is one to determine whether these architectural differences are 
 significant? It is useful to recall that phenomenological theories  
 completely ignore the architecture of the membrane constituents 
 and encapsulate their effects in a small number of parameters which 
 enter the theory, such as the monolayer spontaneous curvature and 
 bending modulus. In that same spirit, we can extract from our simulation 
 those same parameters and compare dimensionless ratios of them to those 
 of other systems. We have done that, and presented the results in Table 
 I. One sees that the values we obtain are reasonable. The ratio of 
 the bilayer compressibility modulus to the hydrophilic/hydrophobic 
 interface tension, $\kappa_a/\gamma_0$,  
 closer to that of liposomes than of polymersomes. The reverse is true 
 for the ratio of the monolayer bending modulus to the product of surface 
 tension and the square of the hydrophobic thickness 
 $\kappa_b/\gamma_0d_c^2$. One line in the table deserves comment, 
that for the 
 experimental values of the bilayer area expansion, $\Delta A/A_0$ quoted 
 at rupture, (the ``critical values"). That for liposomes is smaller than 
 that for polymersomes at rupture, which is equal to the bilayer area 
 expansion we utilized. However the values quoted at 
 rupture have no thermodynamic meaning, because any membrane under tension 
 is inherently unstable and will be observed to rupture if the time scale of 
 observation is sufficiently long. The experimental values quoted apply over 
 some, unspecified, laboratory time scale. On this point we add 
 that, as in experiment, we found many of our bilayers to rupture 
 over the time 
 we observed them, but the time scale for this to happen was significantly 
 greater than that for fusion. If the bilayer area expansion, 
 or equivalently, its tension were reduced, either in experiment or in 
 our simulation, the time scales for the bilayers to fuse and later to 
 rupture would 
 both increase, perhaps to the extent of making impossible 
 the observation of fusion. Indeed we chose the value of tension in the 
 simulation such that fusion could be observed conveniently. 
 One could still ask whether, in addition to increasing 
 the time scale 
 for fusion, a significant reduction in bilayer tension would favor an 
 alternative  
 fusion pathway. To attempt to answer this question, one
 could contemplate even longer Monte Carlo runs on membranes under 
 less tension.  

 There are other physical parameters which might affect the fusion pathway
 but which are not encompassed by the quantities in 
 Table I. For example, one might ask whether the fusion pathway is 
 expected to be the same for large 
 virus-encapsulating endosomes as it is for small synaptic vesicles.
 Thus one would consider the dimensionless 
 ratio of the membrane's hydrophobic thickness to the radius of the 
 vesicle in question. We have considered the simplest case of planar 
 membranes for which this ratio is zero. For endosomes encapsulating
 influenza viruses with an average diameter of $100nm$, the ratio 
 is small, less than, but of the order of, 0.03, 
 but for synaptic vesicles of typical diameter $50nm$, it is at 
 least twice this. It is not difficult to imagine that for a  
 sufficiently large value of this ratio, which implies a small area
 of contact between the fusing vesicles, there might be insufficient room 
 for the growth and movement of the stalk we have observed, so that our 
 mechanism would be supplanted by another. But we do not know this.} 
 \sout{Interestingly, pathway 2, above, is very similar to that
seen independently utilizing a different method, in a very different
system of short rigid amphiphilic rods (Noguchi01).  The apparent
insensitivity to details of the simulation model and methods suggests our
mechanism to be rather generally relevant.}

\begin{figure*}[ht]
\includegraphics[width=\swidth]{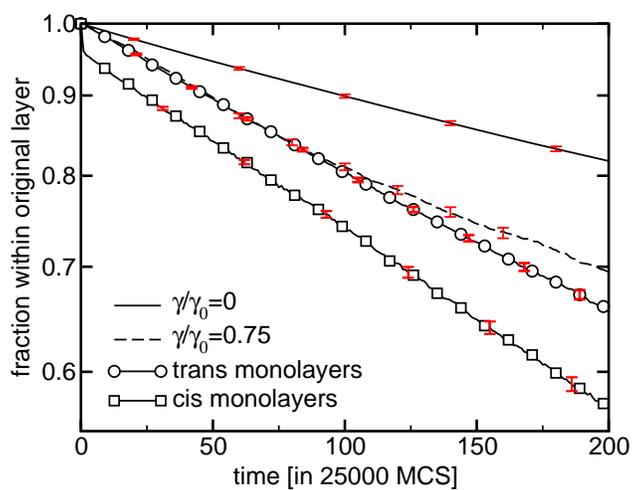}
\caption{Probability of finding an amphiphilic molecule in its original 
monolayer after time $t$.
         The solid and dashed lines refer to simulations of a single 
bilayer under tension 
         $\gamma/\gamma_0=0$ and $\gamma/\gamma_0=0.75$ respectively. 
         The lines with symbols present the results obtained in the simulations
         of fusing bilayers. 
Squares and and circles refer to cis and trans monolayers
         respectively. The time period corresponds to the formation of
stalks and holes. Error bars show standard deviations obtained from
thirty-two  runs.
}
\label{fig:lipid_mix}
\end{figure*}

   Ultimately the most meaningful test of the applicability of our mechanism 
 to biological fusion is comparison to experiment, and our scenario does 
 have 
 testable consequences. First, because of the initial stalk formation, one
expects to see the mixing of lipids in the two proximal layers before the
fusion pore opens, if it forms at all, a result which is in accord with
experiment \cite{Melikyan95,Lee97,Evans02}.
\rev{Second, due to the formation of holes in each bilayer near a stalk,  
our scenario allows for the mixing of those lipids 
in the cis and
trans leaves of one bilayer and also of lipids in the cis leaf of one 
bilayer with those in the trans leaf of the other.  
The standard hemifusion mechanism does not permit either process.}
Note that this movement is different from lipid {\em flip-flop} which is
known to be very slow. \rev{Mixing of lipids between the
cis and trans monolayers has been observed in fusion 
\cite{Lentz97,Evans02}, but it has not yet been determined from which 
membrane they originate and in which membrane they terminate.}  
We have monitored the amphiphiles to see whether they remain in the leaf
in which they were situated at the beginning of the Monte Carlo run, or
mix with amphiphiles in other leaves.  
Instantaneous assignment of amphiphiles to a respective monolayer 
was determined by the center of mass of their hydrophilic part. 
The results are shown in Fig.~\ref{fig:lipid_mix}. \rev{They share with  
experiment the fact that  
the membrane of origin is not distinguished nor is 
the membrane of final residence.}
In order to evaluate the results for the apposed bilayers under tension,
we have
also included those for the single isolated bilayer under zero tension
and under the same tension ($\gamma/\gamma_0=0.75$) as in the simulations of 
fusion. Lateral tension greatly enhances the flip-flop rate in the single
bilayer system. This effect can be explained by an overall thinning of
the membrane, which lowers the translocation barrier, as well as
by the diffusion of amphiphiles through the transient holes formed under
tension. In the simulations of the apposed bilayers, translocation
of amphiphiles from the trans leaves initially follows the same dynamics
as in the single bilayer system, but eventually deviates from it, 
apparently due to the formation of holes facilitated by the
appearance of the stalks, as discussed in the previous section.
Amphiphiles from the cis leaves undergo mixing to the largest extent,
as would be expected due to stalk formation.
Third, our mechanism allows for transient leakage during fusion. 
\rev{As noted earlier, there will be greater leakage if fusion occurs via
pathway two in which the stalk aligns and surrounds two holes than if it
occurs via pathway one in which the stalk rapidly surrounds one hole
before the second appears.} Clearly the amount
of leakage depends on the size of the transient holes formed in the
bilayer, the time between the formation of the initial stalk and the
completion of the fusion pore, and the diffusion constant of the molecules
which leak. 
This constant introduces another time scale whose magnitude, relative to
that of fusion pore formation, determines whether the fusion process is
observed to be leaky or tight.
  
\begin{figure*}[ht]
\includegraphics[width=\swidth]{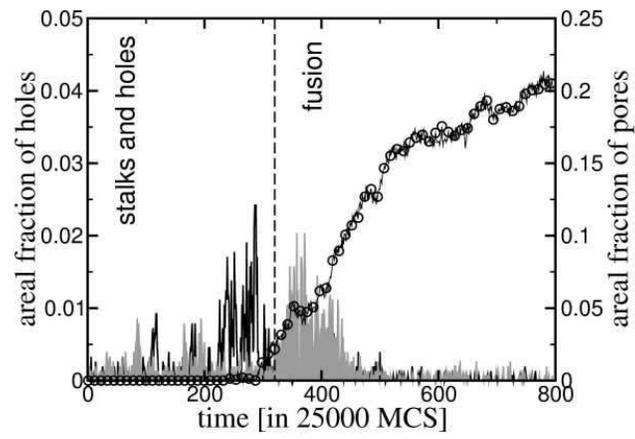}
\caption{Area of pore (symbols) and of holes (lines) vs. time for 
         one simulation run
        (identical to Fig.~\ref{fig:holes_2vs1}). Note the different
        scale for pore and hole areas.
        }
\label{fig:leakage}
\end{figure*}

\rev{It is clear that within our mechanism, leakage via transient holes and
fusion via pore formation are correlated in space and time.
\sout{ as observed experimentally.} The latter
is shown in Fig.~\ref{fig:leakage} which presents, as a function of time, the
area of holes and that of fusion pores from one of the simulation runs.
One sees in this figure, as in the Monte Carlo snapshots,
that the rate at which holes appear, and therefore the rate at which
leakage should occur, {\em increases significantly before, and is correlated
with, the formation of fusion pores}. \sout{Just this effect has been seen in
experiment (Dunina00,FrolovPP)}. Once the fusion pore has
formed, the creation of other holes decreases due to release of
tension initially stored in the membranes.

The question of whether transient leakage is characteristic of 
membrane fusion is an open one. On the one hand, some experiments detect no 
leakage \cite{Spruce91,Tse93,Smit02}, while on the other 
there is a great deal of evidence that
fusion of biological membranes is, indeed, a leaky 
process \cite{Shangguan96,Dunina00,Bonnafous00,Haque02,Smit02}.
It could be
argued that observed leakage is due to the presence, in these
experiments, of fusion
proteins, such as
hemagglutinin, which are certainly present in the vicinity of fusion,
and which are known to undergo conformational changes in which part of
the protein inserts itself into the target vesicle. In support of this
view, one could cite the well known ability of fusion peptides to
initiate erythrocyte hemolysis\cite{Niles90}. Such peptides 
are not included in our model. This argument is vitiated,
however, by the observation that leakage is also detected in the fusion
of model membranes without such peptides 
\cite{Lentz97,Cevc99,Evans02}. In these experiments, large molecules
such as polyethyleneglycol, are used to bring the fusing vesicles
together. It would be difficult to argue that these molecules, which
undergo no conformational change, are responsible for the leakage as they
 generate an attractive 
osmotic force between the vesicles 
precisely because their
large size makes it difficult for them to enter the region where the
vesicles are closely opposed.

One test that might distinguish whether leakage simply accompanies
fusion or is causally related to it is provided by the observation above
that in our mechanism transient leakage is {\em correlated in time and
space} with fusion. Just such an experiment to measure these
correlations has been carried out recently \cite{FrolovPP}, 
and is reported in the paper
accompanying this manuscript. They observe that leakage is, in fact,
correlated spatially and temporally with the process of fusion. Indeed,
their results comparing the time sequence of 
the electrical conductance arising from leakage with that arising from
fusion, shown in their Fig. 5 displays a remarkable similarity to our
results comparing the time sequence of the areal fraction taken up by
holes with that taken up by fusion pores, our Fig. 11.}

%Fourth, our mechanism should exhibit at least two time scales; the first
%for the formation of the fusion pore, and a second  for leakage of contents. 
%The existence of two times
%scales is also consistent with experiment \cite{Evans02}.
%Fifth, 
%there are at least two
%different barriers to overcome for fusion to proceed to completion in
%our mechanism, that for stalk and hole formation, and that for 
%rim formation. As these two processes are uncoupled,
%the leakage rate from holes formed during the
%first process would not be much affected by the second, so that
%the leakage rate would be nearly identical just above and below 
%a threshold for fusion. This is again seen from Fig. 15 which shows that
%hole formation increases before fusion, and decreases after it, but is
%not greatly affected just at the onset of fusion.
%This effect is also observed in experiment \cite{Evans02}.

While the congruence between the predictions of our model and experiment
are very encouraging, there are further tests we should like to apply to
it. Foremost among these is to determine the free energy barriers for the
various steps along the fusion pathway. As noted above, it is relatively
simple to determine the {\em internal energy} during the course of the
simulation as one need only monitor the interactions between
all segments. 
%We have done that, and the results
%are given in Fig.~\ref{fig:energy}. One sees there that the internal energy 
%increases along the fusion pathway by some $100-150 k_BT$. 
But the simulations
cannot easily evaluate the entropy changes along the fusion pathway or,
therefore, the free energy barrier. 
%That the entropy of our fusion
%intermediate is far greater than that of the initial configuration seems
%clear. 
%One need only consider a region of the bilayers that will become
%the site of the stalk, and of its associated holes, and the fusion pore.
%Initially this region fluctuates essentially in one direction, normal to
%the bilayers. The fusion intermediates, however, sample a much larger
%number of configurations because the stalk, and holes, and pores, fluctuate
%in the plane of the bilayers, that is in two dimensions. 
%There is a large entropy gain due to the formation of a highly curved stalk
%intermediate, which permits the hydrophobic portion of the asymmetric
%amphiphiles to sample many more configurations than in a flat monolayer.
%The reason for this comes from the fact that the
%negative spontaneous curvature of the monolayers formed by the asymmetric
%amphiphiles in our system is driven solely by the entropy of chain
%fluctuations, whereas the energy is determined essentially by the
%total hydrophilic-hydrophobic interface. 
%We expect that the large
%change of entropy causes the free energy barrier to be much less than
%the energy barrier. 
To determine the actual value for the free energy 
barrier, calculations using self-consistent field theory, which have
been extremely successful in describing the phase behavior of amphiphiles
\cite{Matsen94,Matsen96} are currently being pursued by us.
Also, elastic constants of the simulated amphiphilic monolayers,
e.g. calculated in \onlinecite{Mueller002}, could be employed
in the simpler phenomenological theories, which have proved to be so useful.
Comparison with the full self-consistent field 
calculations would permit determination of the accuracy of these
elastic models in describing the highly curved intermediates
involved in the fusion reaction.
Furthermore, there is an extensive experimental evidence on
the effect of lipids of differing architecture on fusion 
\cite{Chernomordik96,Zimmerberg99}. 
%We should
%like to be able to determine the effect of such differences in
%architecture on fusion as described by our model. Because
The self-consistent field theory is able to describe such differences 
\cite{Matsen95,Li00} and to
determine both the spatial distribution of different amphiphiles in
inhomogeneous 
structures such as the stalk, the holes and the fusion pore, as
well as the change in the free energy of these structures. 
Results of these investigations will be published separately.

It would  be of great interest to repeat our simulations under
different membrane tension, as this would help to clarify
the importance of fusion peptides in bringing about such tension.
Finally, it would be desirable to carry out simulations in which fusion
peptides are included explicitly. One could investigate whether the membrane
perturbations associated with such inclusions  provide sites for the
nucleation of the small holes that are necessary
for the formation of the fusion pore. If this were so, one
could test the further inference that, by providing 
nucleation sites in close proximity, one in each membrane, such
peptides facilitate sucessful and rapid fusion thereby reducing
leakage.

\begin{acknowledgments}
\vspace{-.8\baselineskip}
We acknowledge very useful conversations with L. Chernomordik, F. Cohen,
M. Kozlov, B. Lentz, D. Siegel, and J. Zimmerberg. We are particularly
grateful to V. Frolov for sharing his knowledge and expertise with us. 
Financial support was provided by the National Science
Foundation under grants  DMR 9876864 and DMR 0140500 and the DFG Bi 314/17 .
Computer time at the NIC J{\"u}lich, the HLR Stuttgart and the 
computing center in Mainz are also gratefully acknowledged. M.M.\
thanks the DFG for a Heisenberg stipend.
\end{acknowledgments}

\clearpage

\bibliography{mmfusion02}
\clearpage

\begin{table}
\begin{tabular}{cccc}
\hline \hline
          & Polymersomes       & Liposomes      & Simulation \\  \hline
$d_c$     & 80\AA              & 30\AA (DOPE$^{(a)}$), 25\AA (DOPC$^{(b)}$)  & 21u        \\ 
$f$       & 0.39               & $0.35\pm0.10$  & 0.34375    \\
$C_0 d_c$ & no data            & -1.1 (DOPE$^{(d)}$), -0.29 (DOPC$^{(c)}$)   & -0.68      \\
$\Delta A/A_0$
          & 0.19               & 0.05           & 0.19       \\
$\kappa_a/\gamma_0$
          & 2.4                & 4.4 (DOPE$^{(b)}$), 2.9 (DOPC$^{(b)}$)   & 4.1        \\
$\kappa_b/\gamma_0 d_c^2$ 
          & 0.044              & 0.10 (DOPE$^{(c)}$), 0.12 (DOPC$^{(d)}$)  & 0.048      \\ 
\hline \hline
\end{tabular}
\caption{Structural and elastic properties of bilayer membranes:
$d_c$ - thickness of membrane hydrophobic core,
$f$ - hydrophilic fraction,
$C_0$ - monolayer spontaneous curvature,
$\Delta A/A_0$ - bilayer area expansion (critical value for the 
experimental systems, and the actual strain used in simulations),
$\kappa_a$ - bilayer area compressibility modulus,
$\kappa_b$ - monolayer bending modulus,
$\gamma_0$ - hydrophilic/hydrophobic interface tension 
\rev{(oil/water tension of 50pN/nm for the experimental systems, 
and A/B homopolymer tension for the simulations).
Data on EO7 polymersomes is taken from \onlinecite{Discher99}; and on lipids
from (a): \onlinecite{Rand89},  (b): \onlinecite{Rand90},
(c): \onlinecite{Chen97}, and (d): \onlinecite{Leikin96}
(see also http://aqueous.labs.brocku.ca/lipid/).
Values of $d_c$, $C_0$ and $\kappa_a$ for DOPE were obtained by linear 
extrapolation from the results on DOPE/DOPC(3:1) mixtures and pure DOPC. 
Values of $\kappa_b$, $\gamma_0$, and $C_0$ for the simulated
model were calculated by us using the method of \onlinecite{Mueller002}}.
}
\label{table1}
\end{table}

\clearpage

%
%\clearpage
%
%
%\clearpage
%
%
%\clearpage
%
%
%\clearpage
%
%
%\clearpage
%
%
%\clearpage
%
%
%\clearpage
%
%
%\clearpage
%
%
%\clearpage
%
%
%\clearpage
%

\end{document}